\documentclass[a4paper,12pt]{article}
\usepackage{txfonts}
\usepackage[utf8x]{inputenc}

\begin{document}

\title{Drude Weight, Meissner Weight, Rotational Inertia of Bosonic
  Superfluids: How Are They Distinguished?}

\author{Bal\'azs Het\'enyi \\ \\
{ \em Department of Physics, Bilkent University} \\ {\em 06800, Ankara, Turkey}
}

\maketitle

\begin{abstract}
{The Drude weight, the quantity which distinguishes metals from
  insulators, is proportional to the second derivative of the ground state
  energy with respect to a flux at zero flux.  The same expression also
  appears in the definition of the Meissner weight, the quantity which
  indicates superconductivity, as well as in the definition of non-classical
  rotational inertia of bosonic superfluids.  It is shown that the difference
  between these quantities depends on the interpretation of the average
  momentum term, which can be understood as the expectation value of the
  total momentum (Drude weight), the sum of the expectation values of single
  momenta (rotational inertia of a superfluid), or the sum over expectation
  values of momentum pairs (Meissner weight).  This distinction appears
  naturally when the current from which the particular transport quantity is
  derived is cast in terms of shift operators. 
}
\end{abstract}

\section{Introduction}

To distinguish conductors from insulators in the quantum case, the strength of
the zero-frequency conductivity was derived by Kohn~\cite{Kohn64}.  The Drude
weight is often expressed~\cite{Kohn64,Shastry90} in terms of the second
derivative of the ground state energy with respect to a phase $\Phi$
associated with the perturbing field as
\begin{equation}
\label{eqn:Dc}
D^{(c)} = \frac{\pi}{V} \left[ \frac{\partial^2 E(\Phi)}{\partial \Phi^2}
  \right]_{\Phi=0},
\end{equation}
where $E(\Phi)$ denotes the perturbed ground state energy.  The Meissner
weight, which appears in London's phenomenological theory of superconductors
is formally identical to the Drude weight.  Moreover, the inverse of the
rotational inertia of a rotating bosonic superfluid (non-classical rotational
inertia (NCRI))is also proportional to the second derivative of the ground state
energy, i.e. it is exactly of the same form as Eq. (\ref{eqn:Dc}), only that
in that case $\Phi$ is proportional to the angular velocity.  Here these
quantities will be collectively called transport susceptibilities.

A fundamental question thus arises: there are three distinct physical
phenomena, but they appear to be described by a single mathematical
expression.  Scalapino, White, and Zhang (SWZ)~\cite{Scalapino92,Scalapino93}
have proposed an interpretation which distinguishes the Drude weight from the
Meissner weight.  They pointed out that the derivative with respect to the
flux is ambiguous.  It could refer to the derivative of the ground state
eigenvalue of the energy with respect to the perturbation (adiabatic
derivative) or the actual ground state as a function of the perturbation
(``envelope'' derivative).  In the absence of level crossings the two are
identical.  SWZ conclude that the difference between the Drude and superfluid
weights is that the former(latter) corresponds to the adiabatic(envelope)
derivative.  Up to now this appears to be the last word on this topic.

There are a number of weaknesses in this interpretation.  In one dimension the
level crossings occur at $\Phi=\pi$ even in the thermodynamic limit, hence in
that case the Drude and Meissner weights can not be distinguished.  Moreover,
as discussed in a recent paper of the author~\cite{Hetenyi12} and summarized
below, the application of these ideas to variational wavefunctions is
ambiguous.  The usual way~\cite{Millis91,Dzierzawa97} to calculate the Drude
weight is to take the second derivative of the variational ground state
energy.  However, this quantity can be cast in terms of an average of the true
energy eigenvalues.  Turning on the perturbation can cause level crossings.
If we insist on SWZ, then these level crossings should be excluded, and the
usual approach~\cite{Millis91,Dzierzawa97} would be invalidated.  In the limit
of a perfect variational wavefunction (one which corresponds to the exact
ground state for any value of the perturbation) the result would be what
according to SWZ is the Meissner weight, not the Drude weight.  We stress
though that this usual approach would {\it only be invalidated if we assume
  SWZ is correct}.  In addition the SWZ interpretation does not distinguish
the Meissner weight associated with superconductors from the non-classical
inertia of rotating superfluids (these two quantities in SWZ belong to the
general category of ``superfluid weight'').

In this paper a different approach to distinguishing the Drude weight,
Meissner weight, and the rotational inertia of a superfluid is developed,
which does not make any reference to whether the derivative is adiabatic or
envelope.  First a general expression for the second derivative of the ground
state energy is derived, which is of the form
\begin{equation}
\label{eqn:Dgen}
\left[ \frac{\partial^2 E(\Phi)}{\partial \Phi^2} \right]_{\Phi=0}= \frac{N}{m} + 
\lim_{\Delta \Phi \rightarrow 0}
\frac{
\langle \Psi | \left[e^{-i\Delta \Phi \hat{X}} \hat{K} + \hat{K} e^{i\Delta
    \Phi \hat{X}} \right] | \Psi \rangle
}{m\Delta \Phi}.
\end{equation}
where $\hat{K}=\sum_{i=1}^N\hat{k}_i$.  In a periodic system this expression
includes an expectation value which can be interpreted in a number of ways.
It can be taken to mean (A) the expectation value of the sum over all momenta,
(B) the sum over expectation values of single momenta, or (C) any other break
up of the total momentum operator (pairs, triplets, etc.).  In fact, this
ambiguity disappears if the current, from which Eq. (\ref{eqn:Dgen}) is
derived, is written in terms of the appropriate Berry phase
expression~\cite{Hetenyi12}.  Case (A) is shown to correspond to the Drude
weight, which distinguishes metallic conductors from insulators.  It is also
shown that metallic conduction can be related to a generalization of the
concept of off-diagonal long-range order (ODLRO)~\cite{Penrose51,Penrose56,Yang62}.  Case (B) is
shown to correspond to bosonic superfluids.  The justification is based on the
fact that the second derivative of the ground state energy with respect to the
flux in this case is proportional to the number of particles in a
Bose-Einstein condensed state.  A direct connection is established between
ODLRO associated with the single particle density matrix.  Breaking up
$\hat{K}$ into pairs is shown to correspond to a condensate of pairs, such as
in the case of BCS superconductivity.  Here, a direct connection is
established between ODLRO associated with the two-particle density matrix.
Moreover, this interpretation, unlike SWZ, distinguishes not only conductors
from superconductors, but also superfluids with single particle condensates
from condensates with other basic groups (two particles, three particles,
etc.).  Also, unlike SWZ, its applicability is independent of dimensions.

The fact that Eq. (\ref{eqn:Dgen}) is ambiguous may appear surprising, but it
can be made obvious by casting the current from which the transport
susceptibility is derived in terms of an explicit position shift
operator~\cite{Hetenyi12}.  In that case, as shown below, the distinct
transport susceptibilities originate from the limiting cases of different
current expressions.

In addition the results of this work solve another open problem.  In a recent
paper Anderson stated~\cite{Anderson12} the following:

\begin{quotation}
{\it ...it has never been demonstrated that ODLRO, and NCRI are synonymous,...}
\end{quotation}
Below this gap is filled by making this connection explicit.

This paper is organized as follows.  The subsequent two sections provide
background information, followed by a brief note on current.  Section
\ref{sec:d2edf2} derives the Drude weight.  In section \ref{sec:ODLRO} the
connection of standard conduction with off-diagonal long-range order is
presented, the subsequent sections treat the case of simple Bose-Einstein
condensation and condensation in a general pairing system.  The penultimate
section presents a comprehensive theory of conduction, after which the work is
concluded.

\section{Background} 

The quantities which in this work will be referred to as transport
susceptibilities are the Drude weight, the Meissner weight (the fraction of
particles which are in a Bose-Einstein condensate in a superconductor), and
the rotational inertia of the superfluid fraction.  In this section some
general background information on transport susceptibilities is provided.

We consider a system of $N$ identical particles in a periodic potential with
Hamiltonian
\begin{equation}
 \hat{H}(\Phi) = \sum_{i=1}^N \frac{(\hat{k}_i + \Phi)^2}{2m} + \hat{V},
\end{equation}
where $\hat{k}_i$ denotes the momentum operator of particle $i$, $m$ denotes
the mass of the particles, $\Phi$ denotes a perturbation, and $\hat{V}$
denotes the interaction potential, for which it holds that
\begin{equation}
V(x_1,...,x_i,...x_N) = V(x_1,...,x_i+L,...x_N)
\end{equation}
for any $i$.  For most of this article, we will consider the ground state of
this Hamiltonian,
\begin{equation}
 \hat{H}(\Phi) | \Psi (\Phi) \rangle = E(\Phi) | \Psi(\Phi) \rangle,
\end{equation}
where $E(\Phi)$($|\Psi(\Phi)\rangle$) denotes the ground state energy
(wavefunction) for the perturbed system.  In the momentum space representation
the unperturbed state can be written as $\Psi(k_1,...,k_N)$, whereas the
perturbed wavefunction takes the form $\Psi(k_1+\Phi,...,k_N+\Phi)$.  One can
also express the pertubed wavefunction in terms of the unpertubed one using
the total momentum shift operator~\cite{Essler05,Hetenyi09} as
\begin{equation}
\label{eqn:Phi_Shift}
 | \Psi(\Phi) \rangle = e^{i\Phi \hat{X}} | \Psi(0) \rangle,
\end{equation}
where $\hat{X} = \sum_{i=1}^N \hat{x}_i$.

The Drude weight was first derived in Ref. \cite{Kohn64}.  The main results
from this work relevant here are that the current and the Drude weight can be
obtained in terms of the first and second derivatives (respectively) of the
ground state energy with respect to $\Phi$, i.e.
\begin{eqnarray}  
J(\Phi) &=& \frac{\partial{E(\Phi)}}{\partial \Phi} \\ \nonumber
D^{(c)} &=& \frac{\pi}{V} \left[ \frac{\partial^2 E(\Phi)}{\partial \Phi^2}
  \right]_{\Phi=0},
\end{eqnarray}  
$D^{(c)}$ is obtained by assuming $\Phi$ to be of the form $\Phi = E
e^{i\omega t}/(i\omega)$.  Using this form for the perturbation, the imaginary
part of the frequency dependent conductivity $\sigma''(\omega)$ can be
calculated and the zero frequency limit of the quantity
$\lim_{\omega\rightarrow0} \omega \sigma''(\omega)$ can be taken, resulting in
$D^{(c)}$.

The Meissner weight is a result of the phenomenological explanation of the
Meissner effect due to London and London.  We follow Ref. \cite{Grosso00}.
We first assume that a superconductor is a perfect conductor, obeying
\begin{equation}
{\bf E} = \frac{1}{n^{(s)}}\frac{\partial {\bf j}}{\partial t},
\end{equation}
where $n^{(s)}$ the density of superconducting charge carriers, ${\bf j}$
indicates the current density.  Using the Maxwell relation for the curl of the
electric field we obtain
\begin{equation}
\frac{\partial}{\partial t} \left[ \nabla\times{\bf j} + n^{(s)}  {\bf B}
  \right]= 0.
\end{equation}
If the quantity in the square brackets is assumed to equal zero then the
Meissner effect can be accounted for and the penetration depth of the magnetic
field in a superconductor can be calculated.  Using this assumption and the London
gauge ($\nabla \chi = 0$) we obtain
\begin{equation}
{\bf j} = n^{(s)} {\bf A}.
\end{equation}
Considering one dimension only, and associating the vector potential with the
momentum shift we obtain
\begin{equation}
n^{(s)} = \frac{1}{V}\left[\frac{\partial^2 E(\Phi)}{\partial \Phi^2}\right]_{\Phi=0}.
\end{equation}

One of the main characteristic properties of a bosonic superfluid emerges from
the rotating bucket experiment, first discussed by Landau~\cite{Landau41} in
1941.  When a superfluid below the critical temperature is rotated slowly, its
moment of inertia is reduced compared to a normal fluid, since the superfluid
fraction remains stationary.  We write the total rotational inertia as
\begin{equation}
I = I^{(s)} + I^{(n)},
\end{equation}
where $I^{(s)}$($I^{(n)}$) corresponds to the rotational inertia associated
with the superfluid(normal) fraction.  Above the critical temperature, where
both fractions rotate, the work to rotate the container would be
\begin{equation}
\Delta W(\Phi) = E(\Phi) - E(0) = I \frac{\Phi^2}{2}.
\label{eqn:ef1}
\end{equation}
Below the critical temperature only the normal fraction would rotate with the
bucket and the work required would be
\begin{equation}
\Delta W^{(n)}(\Phi) = E^{(n)}(\Phi) - E^{(n)}(0) = I^{(n)} \frac{\Phi^2}{2},
\label{eqn:enf1}
\end{equation}
where $E^{(n)}(\Phi)$ denotes the ground state energy associated with the
normal fluid.  From Eqs. (\ref{eqn:ef1}) and (\ref{eqn:enf1}) it follows
that
\begin{equation}
E^{(s)}(\Phi) - E^{(s)}(0) = I^{(s)} \frac{\Phi^2}{2},
\end{equation}
or for small $\Phi$
\begin{equation}
\label{eqn:Is}
I^{(s)} = \left[\frac{\partial^2 E^{(s)}(\Phi)}{\partial \Phi^2}\right]_{\Phi=0}.
\end{equation}
All three quantites $D^{(c)}$, $n^{(s)}$, $I^{(s)}$ are proportional to the
second derivative of the ground state energy with respect to the perturbation
$\Phi$ at $\Phi=0$.

\section{The Problems with Distinguishing Transport 
Susceptibilities Based on  Adiabatic or Envelope Derivatives} 

SWZ suggested~\cite{Scalapino92,Scalapino93} that to distinguish the Drude
weight from the Meissner weight, one has to consider that the derivative with
respect to $\Phi$ in the definition of transport susceptibilities is
ambiguous.  They pointed out that the derivative could refer to the derivative
of the ground state energy with respect to the perturbation (adiabatic
derivative) or that of the zero temperature limit of the free energy (envelope
derivative).  In the case of the former level crossings are excluded.  SWZ
also show that level crossings occur at $\Phi \approx 1/L^{d-1}$, where $L$ is
the linear dimension and $d$ is the dimensionality.  In one dimension the
level crossing occurs at a finite value even in the thermodynamic limit,
resulting in no distinction between the Drude and Meissner weights.  One could
argue that superconductivity is a two-dimensional effect (the Meissner weight
is the response of the system to a magnetic field), but this would be
incorrect.  A superconducting ring is described by a one-dimensional model.
Also the analysis of flux quantization by Byers and Yang~\cite{Byers61} uses a
one-dimensional example (a ring around the cavity).

One can also show that the SWZ interpretation is ambiguous when applied in
variational theory.  The usual procedure to calculate the Drude weight in
variational theory~\cite{Millis91,Dzierzawa97} is to take the second
derivative of the variational ground state energy, however, as shown below,
when this procedure is followed, level crossings are still present, and the
derivative can not be considered adiabatic.  To see this one can compare
variational theory to the finite temperature extension of the Drude weight.

The finite temperature extension of $D^{(c)}$ has been given by Zotos, Castella,
and Prelov\v{s}ek~\cite{Zotos96} (ZCP).  This generalization can be summarized
as
\begin{equation}
D_{adb}(T) = \frac{\pi}{V}\sum_n P_n(0) \left[ \frac{\partial^2 E_n(\Phi)}{\partial
    \Phi^2} \right]_{\Phi=0},
\label{eqn:ZCP_Dc}
\end{equation}
where
\begin{equation}
P_n(0) = \frac{\exp\left(-\frac{E_n(0)}{k_B T}\right)}{Q(0)},
\end{equation}
and where $Q(0)$ denotes the canonical partition function of the unperturbed
system.  The important point is that in Eq. (\ref{eqn:ZCP_Dc}) the Boltzmann
weight factors {\it remain unchanged as the perturbation $\Phi$ is turned on.}
Thus the effect of level crossings is excluded and the {\it derivative is the
  adiabatic one.}  Taking the zero temperature limit reproduces Kohn's
expression for $D$ (Eq. (\ref{eqn:Dc})).  Eq. (\ref{eqn:ZCP_Dc}) consists of a
sum over {\it adiabatic derivatives} of energies weighted by the Boltzmann
factor.  Eq. (\ref{eqn:ZCP_Dc}) has been applied~\cite{Kirchner99} to
calculate the Drude weight in strongly correlated systems.

To define~\cite{Hetenyi12} a quantity which in the limit of zero temperature
produces Eq. (\ref{eqn:Dc}), but with the envelope derivative instead of the
adiabatic one, one could modify Eq. (\ref{eqn:ZCP_Dc}) as
\begin{equation}
D_{env}(T) = \frac{\pi}{V} \left[ \frac{\partial^2}{\partial \Phi^2} \langle E(\Phi)
\rangle \right]_{\Phi=0} = \frac{\pi}{V}
  \frac{\partial^2}{\partial \Phi^2} \left[ \sum_n P_n(\Phi) E_n(\Phi)
    \right]_{\Phi=0},
\label{eqn:ZCP_Ds}
\end{equation}
where $\langle E(\Phi) \rangle$ indicates the average energy of the perturbed
system.  Alternatively, one could also define a quantity based on the free
energy as
\begin{equation}
\label{eqn:ZCP_Ds2}
D_{env}(T) = \frac{\pi}{V} \left[ \frac{\partial^2 F(\Phi)}{\partial \Phi^2}
  \right]_{\Phi=0}.
\end{equation}
In the zero temperature limit both Eqs. (\ref{eqn:ZCP_Ds}) and
(\ref{eqn:ZCP_Ds2}) tend to the same expression, Eq. (\ref{eqn:Dc}) but this
time with the {\it envelope derivative,} since level crossings can in this
case alter the state which enters the definition of the derivative.

In a variational theory, when the Drude weight is calculated,
usually~\cite{Millis91,Dzierzawa97} the second derivative of the variational
energy is taken with respect to $\Phi$ .  Such an assumption is not consistent
with the SWZ interpretation for the following reasons.  Suppose
$|\tilde{\Psi}(\gamma)\rangle$ is a variational wavefunction, where $\gamma$
denotes a set of variational parameters, which we wish to use to optimize some
Hamiltonian $\hat{H}$ with eigenbasis
\begin{equation}
\hat{H}|\Psi_n \rangle = E_n |\Psi_n \rangle.
\end{equation}
The estimate for the ground state energy may be written in terms of a density
matrix as
\begin{equation}
\langle \tilde{\Psi}(\gamma)|\hat{H}|\tilde{\Psi}(\gamma)\rangle = \sum_n
\langle \tilde{\Psi}(\gamma)|\Psi_n\rangle E_n \langle \Psi_n
|\tilde{\Psi}(\gamma)\rangle = \sum_n \tilde{P}_n E_n,
\end{equation}
the probabilities can be written as
\begin{equation}
\tilde{P}_n = |\langle \tilde{\Psi}(\gamma) | \Psi_n \rangle|^2.
\end{equation}
Comparing with Eq. (\ref{eqn:ZCP_Dc}) it is obvious that if the SWZ
interpretation is assumed then the correct Drude weight would be defined as
\begin{equation}
\label{eqn:Dcv}
D_{adb} = \sum_n \tilde{P}_n(0) \left[ \frac{\partial^2 E_n(\Phi)}{\partial
    \Phi^2} \right]_{\Phi=0},
\end{equation}
with $\tilde{P}_n(0)$ independent of the perturbation $\Phi$, since this way
we would have a set of weighted adiabatic derivatives.  In this case the
effect of level crossings on the weights would be excluded.

Instead, the standard way~\cite{Millis91,Dzierzawa97} to calculate the Drude
weight in variational theory is to take the second derivative of the
variational energy with respect to the perturbation, i.e.
\begin{equation}
D_{env} = \frac{\pi}{V}\frac{\partial^2 }{\partial \Phi^2} \left[\sum_n P_n(\Phi) E_n(\Phi)\right]_{\Phi=0},
\label{eqn:Dsv}
\end{equation}
in which case the effect of level crossings are {\it not excluded,} and which
corresponds to an {\it envelope derivative.}  In fact Eq. (\ref{eqn:Dsv}) has
the same form as Eq. (\ref{eqn:ZCP_Ds}), in both cases the derivatives of the
average energy are taken.  Millis and Coppersmith~\cite{Millis91} conclude
based on Eq. (\ref{eqn:Dsv}) that the Gutzwiller projected Fermi
sea~\cite{Gutzwiller65} is a conductor.  The equivalent of the zero
temperature limit for Eq. (\ref{eqn:Dsv}) would be the limit of a perfect
variational wavefunction, which corresponds to the true wavefunction for any
value of $\Phi$.  In this limit Eq. (\ref{eqn:Dsv}) would corresponds to the
{\it envelope derivative}, in other words, according to the logic of SWZ, the
Meissner weight.  It needs to be stressed that the statement of this article
is not that Eq. (\ref{eqn:Dsv}) corresponds to the Meissner weight, only that
it does according to the criteria of SWZ.

Apart from the above, another shortcoming of the SWZ prescription is that it
does not explicitly distinguish bosonic superfluids from superconductors
(condensation of paired fermions). 

\section{Berry Phase Expression for the Current in Many-Body Systems with
  Periodic Boundary Conditions}

In Ref. \cite{Hetenyi12} it was shown that for continuous systems with
many-particles under periodic boundary conditions the current can be expressed
as
\begin{equation}
\label{eqn:JBP_DX}
J_N(\Phi) = \frac{N}{m}\Phi + \lim_{\Delta X \rightarrow 0} \frac{1}{m\Delta X}
\mbox{Im} \ln \langle \Psi(\Phi) | \exp(i \Delta X \hat{K})|\Psi(\Phi) \rangle.
\end{equation}
Carrying out the limit $\Delta X \rightarrow 0$ results in
\begin{equation}
\label{eqn:JBP_DX_lim}
J_N(\Phi) = \frac{N}{m}\Phi +  \frac{1}{m} \langle \Psi(\Phi) | \hat{K}|\Psi(\Phi) \rangle.
\end{equation}
However, for a system with identical particles one could also write 
\begin{equation}
\label{eqn:JBP_DX1}
J_1(\Phi) = \frac{N}{m}\Phi + \lim_{\Delta X \rightarrow 0} \frac{N}{m\Delta X}
\mbox{Im} \ln \langle \Psi(\Phi) | \exp(i \Delta X \hat{k})|\Psi(\Phi) \rangle,
\end{equation}
where $\hat{k}$ is a single momentum operator, or more generally one has
\begin{equation}
\label{eqn:JBP_DXp}
J_p(\Phi) = \frac{N}{m}\Phi + \lim_{\Delta X \rightarrow 0} \frac{N/p}{m\Delta X}
\mbox{Im} \ln \langle \Psi(\Phi) | \exp\left(i \Delta X \sum_{i=1}^p \hat{k}_i\right)|\Psi(\Phi) \rangle.
\end{equation}
Carrying out the limit in $\Delta X \rightarrow 0$ Eqs. (\ref{eqn:JBP_DXp})
and (\ref{eqn:JBP_DX1}) would appear to give identical results similar to
Eq. (\ref{eqn:JBP_DX_lim}).

The difference between $J_p(\Phi)$ for different $p$s becomes obvious if we
cast the second term in terms of the appropriate reduced density matrix
(Eq. (\ref{eqn:rdm})),
\begin{equation}
\label{eqn:JBP_DXp_rdm}
J_p(\Phi) = \frac{N}{m}\Phi + \lim_{\Delta X \rightarrow 0} \frac{N/p}{m\Delta
  X} \mbox{Im} \ln \mbox{Tr} \left\{\hat{\rho}_p\exp\left(i \Delta X
\sum_{i=1}^p \hat{k}_i\right)\right\}.
\end{equation}
As shown below the transport susceptibilities derived from a particular
definition of current are sensitive to ODLRO~\cite{Yang62} in density matrices
of different orders.  In the examples analyzed below, it will always be
assumed that the transport susceptibility is derived from one particular
definition of the current, Eq. (\ref{eqn:JBP_DXp_rdm}), i.e., a particular
value of $p$.  However, to prevent the notation from becoming too cumbersome,
we will not write the current in terms of the corresponding shift operators.

\section{Expressing $\left[\frac{\partial^2 E(\Phi)}{\partial
      \Phi^2}\right]_{\Phi=0}$}

\label{sec:d2edf2}

As our first example we analyze the case $p=N$.  The quantities in this
section are derived based on $J_N(\Phi)$.  The first derivative of the ground
state energy with respect to $\Phi$ corresponds to the total current, and,
after the limit $\Delta X \rightarrow 0$ is taken, it can be written as
\begin{equation}
\label{eqn:J}
 J(\Phi) = \frac{N}{m}\Phi + \frac{\langle \Psi(\Phi) | \hat{K} | \Psi(\Phi) \rangle}{m}.
\end{equation}
Taking the next derivative results in
\begin{equation}
 \frac{\partial^2 E(\Phi)}{\partial \Phi^2} = \frac{N}{m} 
+ \frac{1}{m} \left[
   \langle \partial_\Phi \Psi(\Phi) | \hat{K} | \Psi(\Phi) \rangle + \langle
   \Psi(\Phi) | \hat{K} | \partial_\Phi\Psi(\Phi) \rangle \right].
\end{equation}
We now multiply and divide the last two terms by $\Delta \Phi$, resulting in
\begin{equation}
 \frac{\partial^2 E(\Phi)}{\partial \Phi^2} = \frac{N}{m} 
+ \frac{1}{m\Delta \Phi} \left[
   \Delta \Phi \langle \partial_\Phi \Psi(\Phi) | \hat{K} | \Psi(\Phi) \rangle
   + \Delta \Phi \langle \Psi(\Phi) | \hat{K} | \partial_\Phi\Psi(\Phi) \rangle \right].
\end{equation}
In the limit $\Delta \Phi \rightarrow 0$ it holds that
\begin{equation}
\Delta \Phi \langle \partial_\Phi \Psi(\Phi) | = 
\langle \Psi(\Phi + \Delta \Phi) | - 
\langle \Psi(\Phi ) |. 
\end{equation}
Using the fact that at $\Phi=0$ the total current is zero, we obtain
\begin{equation}
 \frac{\partial^2 E(\Phi)}{\partial \Phi^2} = \frac{N}{m} 
+ \lim_{\Delta \Phi \rightarrow 0}\frac{1}{m\Delta \Phi} \left[
    \langle  \Psi(\Delta \Phi) | \hat{K} | \Psi(0) \rangle
   +  \langle \Psi(0) | \hat{K} | \Psi(\Delta \Phi) \rangle \right].
\end{equation}
Applying the definition of the shift operator results in
\begin{equation}
\label{eqn:Dgen2}
 \frac{\partial^2 E(\Phi)}{\partial \Phi^2} = \frac{N}{m} 
+ \lim_{\Delta \Phi \rightarrow 0}\frac{1}{m\Delta \Phi} \left[
    \langle  \Psi | e^{-i\Delta \Phi \hat{X}}\hat{K} | \Psi \rangle
   +  \langle \Psi | \hat{K} e^{i\Delta \Phi \hat{X}}| \Psi \rangle \right].
\end{equation}
which is the same as Eq. (\ref{eqn:Dgen}).

The interpretation of Eq. (\ref{eqn:Dgen2}) is the same as that of the Drude
weight derived in Ref. \cite{Hetenyi13}.  If the unperturbed wavefunction
$|\Psi\rangle$ is an eigenstate of $\hat{K}$, given that it is unperturbed it
would have to have an eigenvalue of zero.  In this case the second derivative
is simply $\frac{N}{m}$.  When that is not the case one can expand
Eq. (\ref{eqn:Dgen}) in $\Delta \Phi$ and keep the leading term, resulting in
\begin{equation}
\left[ \frac{\partial^2 E(\Phi)}{\partial \Phi^2} \right]_{\Phi=0}= \frac{N}{m} + 
i\frac{
\langle \Psi(0) | [\hat{K},\hat{X}] | \Psi(0) \rangle
}{m}.
\end{equation}
The zeroth order term in the expansion in $\Delta \Phi$ corresponds to the
expectation value of the total current in the unperturbed state which is zero.
Using the definitons of the operators $\hat{K}$ and $\hat{X}$, it is easy to
show that
\begin{equation}
 [\hat{K},\hat{X}] = \sum_{i=1}^N [\hat{k_i},\hat{x}_i] = i N,
\end{equation}
and that the second derivative in this case is zero.

\section{Off-Diagonal Long-Range Order}  

\label{sec:ODLRO}

One can also cast~\cite{Hetenyi12a} the criterion for conduction in terms
discontinuous features of the distribution of the total momentum $K$
alternatively in terms of a variation on the idea of ODLRO.  We define
\begin{equation}
 P_N(K) = \int...\int dk_1...dk_N |\Psi(k_1,...,k_N)|^2
 \delta\left(K-\sum_{i=1}^N k_i\right).
\end{equation}
If $\Psi$ is an eigenstate of the total momentum, $P_N(K)$ is a $\delta$-peak
at the origin.  For the insulating state $P_N(K)$ is some smooth function,
symmetric around the origin.  One can define a quantity,
\begin{equation}
 \tilde{\rho}_N(X,X') = \int...\int dx_1...dx_N  \Psi(x_1+X,...,x_N+X) \Psi^*(x_1+X',...,x_N+X').
\end{equation}
It is easy to show that
\begin{equation}
 \tilde{\rho}_N(X,X') = \int dK P_N(K) e^{iK(X-X')},
\end{equation}
and that conduction corresponds to
\begin{equation}
 \lim_{|X-X'|\rightarrow \infty }\tilde{\rho}_N(X,X') = \mbox{finite},
\end{equation}
whereas insulation corresponds to a decay in $\tilde{\rho}_N(X,X')$ to zero.

In the following we will use the reduced density matrices defined as
\begin{equation}
\label{eqn:rdm}
\rho_p(x_1,...,x_p;x'_1,...,x'_p) = \int...\int dx_{p+1}...dx_N 
  \Psi(x_1,...,x_p,x_{p+1},...,x_N)
 \Psi(x'_1,...,x'_p,x_{p+1},...,x_N).
\end{equation}
It is well-known~\cite{Yang62} that long-range order in the reduced density
matrix corresponds to Bose-Einstein condensation in systems of identical
particles at low temperature.  For example, if the one-body reduced density
matrix exhibits long-range order, i.e.
\begin{equation}
 \lim_{|x-x'|\rightarrow \infty}\rho_1(x;x') = \mbox{finite},
\end{equation}
then the system exhibits condensation in which the basic group has one
particle (superfluidity in bosonic systems, e.g. He$^4$).  Similarly, ODLRO in
$\rho_2(x_1,x_2;x'_1,x'_2)$, but not in $\rho_1(x;x')$ corresponds to the
condensation where the basic group consists of two particles, as in BCS
pairing in superconductors, or superfluidity in He$^3$.  ODLRO in the $m$-body
real-space reduced density matrices corresponds to $\delta$-peaks in the
$m$-body momentum distributions.  These results were derived by Yang
\cite{Yang62}.  Yang has also shown that if off-diagonal long-range order is
present in some reduced density matrix $\rho_j$, then it will also be present
in all reduced density matrices $\rho_k$ with $k\geq j$.

\section{Bose-Einstein Condensation of Single Particles}  

We will now interpret Eq. (\ref{eqn:Dgen}) as a sum over single-particle 
 momenta, in
other words, we assume that the current expression from which the transport
susceptibility originates is $J_1(\Phi)$ (Eq. (\ref{eqn:JBP_DX1})).  The
corresponding second derivative is
\begin{equation}
\label{eqn:Dc3}
\left[ \frac{\partial^2 E(\Phi)}{\partial \Phi^2} \right]_{\Phi=0}= \frac{N}{m} + 
\sum_{j=1}^N \lim_{\Delta \Phi \rightarrow 0}
\frac{
\langle \Psi | \left[e^{-i\Delta \Phi \hat{X}} \hat{k}_j  + 
 \hat{k}_j e^{i\Delta \Phi \hat{X}}\right]| \Psi \rangle
}{m\Delta \Phi}.
\end{equation}
Equations (\ref{eqn:Dgen}) and (\ref{eqn:Dc3}) appear to be identical, however
they are distinct, with different physical meanings.  We first expand in
$\Delta \Phi$ resulting in
\begin{equation}
\label{eqn:Ds}
\left[ \frac{\partial^2 E(\Phi)}{\partial \Phi^2} \right]_{\Phi=0}= \frac{N}{m} + 
i\sum_{j=1}^N\frac{
\langle \Psi | [\hat{k}_j,\hat{x}_j] | \Psi \rangle
}{m}.
\end{equation}
The second part of Eq. (\ref{eqn:Ds}) is an average over single particle
commutators.  This average can be expressed in terms of the one body reduced
density matrix as
\begin{equation}
\label{eqn:Dsrdm}
 i\sum_{j=1}^N\frac{
\langle \Psi | [\hat{k}_j,\hat{x}_j] | \Psi \rangle
}{m} = \frac{iN}{m} \mbox{Tr} \hat{\rho}_1 [\hat{k},\hat{x}].
\end{equation}
The one-body reduced density matrix can be diagonalized resulting in
\begin{equation}
 \rho_1(x;x') = \sum_j R^{(1)}_j f_j(x) f_j(x'),
\end{equation}
where $f_j(x)$ are the natural orbitals of the many-body system,
\begin{equation}
 \sum_j R^{(1)}_j = 1,
\end{equation}
and $R^{(1)}_j \geq 0$ for all $j$.  In order to evaluate
Eq. (\ref{eqn:Dsrdm}) we first consider the action of the commutator on a
single orbital.  In general it will hold that
\begin{equation}
\langle f_j |[\hat{k},\hat{x}]| f_j \rangle = i ,
\end{equation}
except if $f_j(x)$ is an eigenstate of either the momentum or the
position.~\cite{CommProof}  In particular for the zero momentum state
\begin{equation}
 f_j(x) = \frac{1}{\sqrt{V}},
\end{equation}
it holds that
\begin{equation}
\label{eqn:crit01}
 \langle f_j |[\hat{k},\hat{x}]| f_j \rangle = 0.
\end{equation}
Such eigenstates of the reduced density matrix will not contribute to the
average in the second term in Eq. (\ref{eqn:Ds}) so
\begin{equation}
 \left[ \frac{\partial^2 E(\Phi)}{\partial \Phi^2}
   \right]_{\Phi=0}=\frac{N}{m}\left(1-\sum_j' R_j \right)
=\frac{N_0}{m},
\end{equation}
where the prime indicates that the summation is over states which are not zero
momentum states.  $N_0\leq N$ can be associated with the number of particles
in zero momentum states.  In principle it can also occur that an
  eigenstate of the reduced density matrix is also an eigenstate of the
  momentum, but with a finite eigenvalue.  Such states will contribute to the
  non-classical rotational inertia, but not to ODLRO.  In this sense
  Bose-Einstein condensation is distinct from superfluidity.
 
Clearly, the expression for the second derivative of the energy, when
interpreted according to Eq. (\ref{eqn:Dc3}), is proportional to the number of
single particles in a zero momentum state, in other words the Bose-Einstein
condensate, therefore we can interpret the second derivative in this case as
the rotational inertia of the superfluid component of a rotating sample
($I^{(s)}$ Eq. (\ref{eqn:Is})).  Also, the casting of Eq. (\ref{eqn:Dgen}) in
terms of the one-body reduced density matrix establishes the connection
between non-classical rotational inertia of a superfluid and off-diagonal long
range order, solving a long-standing open problem~\cite{Anderson12}.

\section{Bose-Einstein Condensation of Pairs of Particles}  

One can also break up the total momentum operator into pairs of momenta,
rather than only single momenta.  We will use first quantization, as we have
throughout the paper.  Some details of the first quantized notation in the
context of indistinguishable particles is given in the appendix.

In this case the current from which the transport susceptibility is derived is
of the form $J_2(\Phi)$ (Eq. (\ref{eqn:JBP_DXp}) with $p=2$).  The second
derivative of the energy, when the current is taken to mean a sum over pairs,
takes the form
\begin{equation}
\label{eqn:Dsc1}
\left[ \frac{\partial^2 E(\Phi)}{\partial \Phi^2} \right]_{\Phi=0}= \frac{N}{m} + 
\sum_{j=1}^{\frac{N}{2}} \lim_{\Delta \Phi \rightarrow 0}
\frac{
\langle \Psi | \left[ e^{-i\Delta \Phi \hat{X}} \hat{k}^{(2)}_{j}  +
  \hat{k}^{(2)}_j e^{i\Delta \Phi \hat{X}}\right] | \Psi \rangle
}{m\Delta \Phi},
\end{equation}
where
\begin{equation}
\label{eqn:k2}
 \hat{k}^{(2)}_j = \hat{k}_j + \hat{k}_{j+\frac{N}{2}}.
\end{equation}
Note that the indices on operators refer to arguments of the wavefunction on
which $\hat{k}_j + \hat{k}_{j+\frac{N}{2}}$ operates.  Taking the limit
$\Delta \Phi \rightarrow 0$ leads to
\begin{equation}
\label{eqn:Dsc2}
\left[ \frac{\partial^2 E(\Phi)}{\partial \Phi^2} \right]_{\Phi=0}= \frac{N}{m} + 
i\sum_{j=1}^{\frac{N}{2}}\frac{
\langle \Psi | [\hat{k}^{(2)}_j,\hat{x}^{(2)}_j] | \Psi \rangle
}{m}.
\end{equation}
Due to the indistinguishability of the particles we can cast
Eq. (\ref{eqn:Dsc2}) in terms of the two-body reduced density matrix,
\begin{equation}
\label{eqn:Dsc3}
\left[ \frac{\partial^2 E(\Phi)}{\partial \Phi^2} \right]_{\Phi=0}= \frac{N}{m} + 
\frac{iN}{2m}\mbox{Tr} \{\hat{\rho}_2 [\hat{k}^{(2)},\hat{x}^{(2)}]\}.
\end{equation}
As before we can diagonalize $\hat{\rho}_2$ as
\begin{equation}
 \rho_2(x_1,x_2;x'_1,x'_2) = \sum_j R^{(2)}_j g_j(x_1,x_2)g_j(x'_1,x'_2),
\end{equation}
where
\begin{equation}
 \sum_j R^{(2)}_j = 1,
\end{equation}
with $R^{(2)}_j\geq0$ for all $j$.  In general it holds that
\begin{equation}
\label{eqn:id_com}
[\hat{k}^{(2)},\hat{x}^{(2)}] g_j(x_1,x_2) = 2 i g_j(x_1,x_2),
\end{equation}
except for pair-orbitals $g_j(x_1,x_2)$ for which
\begin{equation}
 \hat{k}^{(2)} g_j(x_1,x_2) = 0.
\end{equation}
(This would be the case for BCS pairs, since there the momenta of opposite
spin particles cancel.)  Again, such pairing states will not contribute to the
second derivative of the energy, since
\begin{equation}
\langle g_j |[\hat{k}^{(2)},\hat{x}^{(2)}]| g_j \rangle  = 0,
\end{equation}
whereas for the rest we can use Eq. (\ref{eqn:id_com}), resulting in
\begin{equation}
 \left[ \frac{\partial^2 E(\Phi)}{\partial \Phi^2} \right]_{\Phi=0}=
 \frac{N(1-\sum'_{j=1}R^{(2)}_j)}{m} = \frac{N_{0,p}}{m}.
\end{equation}
$N_{0,p}\leq N$ can be interpreted as the number of electrons in paired states
for which the total momentum is zero, such as Cooper pairs in the BCS theory.

\section{A Comprehensive Theory of Transport}  
\label{sec:cmpthr}

Based on the above one can define a generalized transport susceptibility as
follows:
\begin{equation}
\label{eqn:Dp}
D_p = \frac{\pi}{V} \left[ \frac{\partial J_p(\Phi)}{ \partial \Phi }
  \right]_{\Phi=0} =
\frac{\pi}{V}\left( \frac{N}{m} + 
\sum_{j=1}^{\frac{N}{p}} \lim_{\Delta \Phi \rightarrow 0}
\frac{
\langle \Psi | \left[ e^{-i\Delta \Phi \hat{X}} \hat{k}^{(p)}_{j} + \hat{k}^{(p)}_j e^{i\Delta \Phi \hat{X}}\right]| \Psi \rangle
}{m\Delta \Phi} \right),
\end{equation}
where $\hat{k}^{(p)}_{j}$ indicates the sum of $p$ distinct momenta.  The
quantity $D_p$ can be used to distinguish insulators and different types of
conductors.  Due to the result of Yang~\cite{Yang62}, that if ODLRO is present
in a reduced density matrix of order $p$, then all higher order density
matrices will also exhibit ODLRO, it follows that if $D_p$ is finite, then all
$D_r$'s will be finite if $r\geq p$.  Hence, in principle, one has to find the
smallest value $p_m$ for which $D_{p_m}$ is finite.  If $p_m$ is of
microscopic magnitude ($p_m=1$ for bosonic superfluids or perfect
conductors, $p_m=2$ for BCS superconductors) then the system can
be classified as a superconductor.  If $p_m$ is on the order of the total
number of particles in the system, then the system can be classified as a
regular conductor.  If all $D_p=0$ then the system is an insulator.

We have defined a large number of $D_p$, which raises the question: which one
is measured experimentally?  Experiments detect the motion of particles,
charges in conductors, so a finite $D_p$, for any value of $p$ will be
detected as conduction.  To decide whether the current corresponds to a
Bose-Einstein condensed state (bosonic superfluids, superconductors),
information other than conduction is needed (for example Meissner effect, flux
quantization, non-classical rotational inertia).

Another interesting aspect of the above results is the overall interpretation
of conductivity which follows.  Bose-Einstein condensates are independent
particles in zero momentum states.  Superconductors are pairs of particles in
zero momentum states.  Normal conduction in a correlated system is a large
(thermodynamic) number of particles in zero momentum states.  In a
superconductor the applied field moves pairs of particles independently,
whereas in a normal conductor, a large number of particles are moved together.
This is consistent with the fact that a superconductor can sustain a
persistent current for a very long time, as well as with Kohn's
theory~\cite{Kohn64} of normal conductors.  Kohn's statement is that
insulation is a result of many-body localization, the wavefunction is a linear
combination of states each of which includes large number of particles which
are localized.  An equivalent statement is: a conducting state is one in which
a large number of particles are simultaneously delocalized~\cite{Hetenyi13}.

One more aspect of the above needs to be mentioned.  It turns out that $D_1$
is finite for a Fermi sea, since the wavefunction in this case consists of a
Slater determinant of eigenstates of the single particle momentum operator
(non-interacting system).  If an interaction is turned on, however small (for
example the case of a Landau Fermi liquid), then one expects that the
wavefunction will not consist of a Slater determinant of eigenstates of the
single momentum, and $D_1$ will, in general, not be finite.  However the fact
that $D_1$ is finite suggests that the Fermi sea exhibits properties similar
to bosonic superfluids.  It has recently been suggested by
Hirsch~\cite{Hirsch13,Hirsch09} that current in a superconductor is carried by free
electrons.

The evaluation of a particular $D_p$ consists of the following steps.  First
the reduced density matrix of order $p$ is calculated, diagonalized, and its
eigenstates obtained.  Then, for each state it needs to be determined whether
it is an eigenstate of the $p$-body momentum operator, $\hat{K}_p =
\sum_{i=1}^p \hat{k}_i$.  This operator when applied directly to the state
will reduce to a sum of single body momenta, so it is essential to use the
operator $\exp(i\Delta X \hat{K}_p)$ (which is a partial shift operator).  If
a particular eigenstate of the reduced density matrix is also an eigenstate of
$\exp(i\Delta X \hat{K}_p)$, then it contributes to $D_p$, otherwise it does
not.

On the technical side the main issue is the evaluation and subsequent
diagonalization of the reduced density matrix.  This already has a history,
since it is an important step also in the study of natural
orbitals~\cite{Davidson72}, and more recently in the density matrix
renormalization group method.~\cite{Schollwoeck05} These statements are valid
for the calculation of actual models, as well as variational theories.  In the
latter case the first step is the calculation of the reduced density matrix
associated with the variational wavefunction.

\section{Conclusion}  

In this paper the problem of transport susceptibilities was considered.  It is
well-known that the Drude weight, the Meissner weight, and the rotational
inertia of a rotating superfluid all have the same mathematical expressions
apart from constants factors.  This problem was thought to have been solved by
Scalapino, White, and Zhang, based on observing an ambiguity in the definition
of the derivative of the ground state energy, namely, that the derivative
could refer to the adiabatic or the envelope derivatives.  This classification
is not applicable in one dimension, is cumbersome to apply consistently in a
variational setting, and only divides the transport susceptibilities into two
categories (conductor and superfluid).

In this paper, it was shown that a more fruitful approach to the problem is to
start with the Berry phase expression for the current, and distinguish between
currents in which the charge carriers conduct individually, in pairs, or in
larger (thermodynamically large) clusters.  A particular current and the
susceptibility derived from it can be cast in terms of a reduced density
matrix of the corresponding order (order one for the individually conducting
case, order two for paired systems, etc.), and its value will be sensitive to
off-diagonal long range order in the reduced density matrix of the given
order.  Thus the susceptibility for the case of conduction by
thermodynamically large clusters corresponds to the Drude weight, for the
paired case to the Meissner weight.  For the case of individual particles the
non-classical rotational inertia results.

\section*{Acknowledgments}{ The author is grateful to the Physical Society of
Japan for financial support in publication.  The author acknowledges financial
support from the Turkish agency for basic research (T\"UBITAK, grants
no. 112T176 and 113F334).}

\appendix
\section{Anti-Symmetry and First Quantization}

In Eq. (\ref{eqn:k2}) a two-body operator is defined in first quantization
whose expectation value is subsequently evaluated over a many-body
wavefunction of indistinguishable particles.  In this appendix a brief
discussion of first-quantized operators in the context of indistinguishable
particles is presented, for a more complete discussion the reader may consult
Ref. \cite{Reichl09}.

The indices in Eq. (\ref{eqn:k2}) refer to the positions of arguments in the
wavefunction.  To give an example, let us consider a three-particle system of
spinless fermions in the state
\begin{equation}
|\Psi(k_1,k_2,k_3)\rangle = \frac{1}{\sqrt{3!}}\left(
|k_1,k_2,k_3\rangle - |k_1,k_3,k_2\rangle  
 -|k_2,k_1,k_3\rangle + |k_2,k_3,k_1\rangle  
+|k_3,k_1,k_2\rangle - |k_3,k_2,k_1\rangle  
\right).
\end{equation}
$\Psi(k_1,k_2,k_3)$ is an antisymmetric wavefunction, hence it is a valid
wavefunction for three identical fermions.

One can define an operator
\begin{equation}
 \hat{k}^{(2)} = \hat{k}_1 + \hat{k}_2,
\end{equation}
which when it acts on one of the (not antisymmetric) components of $\Psi$
results in the sum of the values of the momenta in the first and second
arguments, for example,    
\begin{equation}
 \hat{k}^{(2)}|k_3,k_2,k_1\rangle =  (k_3 + k_2)|k_3,k_2,k_1\rangle
\end{equation}
or
\begin{equation}
 \hat{k}^{(2)}|k_1,k_3,k_2\rangle =  (k_1 + k_3)|k_3,k_2,k_1\rangle.
\end{equation}
Using this one can easily show that
\begin{equation}
\langle \Psi |  \hat{k}^{(2)} | \Psi \rangle = 2 \bar{k},
\end{equation}
where 
\begin{equation}
\bar{k} = \frac{k_1+k_1+k_3}{3},
\end{equation}
in other words the average momentum.


\begin{thebibliography}{9}
\bibitem{CommProof}  For any two operators $\hat{A}$ and $\hat{B}$, the
  expectation value $\langle \phi_j | [\hat{A},\hat{B}] |\phi_j \rangle = 0$
  if $\hat{A} | \phi_j \rangle = A_j | \phi_j \rangle $, in other words if $|
  \phi_j \rangle $ is an eigenstate of $\hat{A}$.
\end{thebibliography}
\end{document}